\documentclass[sort&compress,cmfonts
    ,final            
  ]
  {aipproc}

\layoutstyle{6x9}

\begin{document}
\newcommand{\lsim}{\, \lower2truept\hbox{${<\atop\hbox{\raise4truept
\hbox{$\sim$}}$}}\,}
\newcommand{\gsim}{\,
\lower2truept\hbox{${>\atop\hbox{\raise4truept\hbox{$\sim$}}$}}\,}

\title{The Sunyaev-Zeldovich effect as a probe of the galaxy formation process}


\author{G. De Zotti}{
  address={INAF, Osservatorio Astronomico di Padova, Vicolo dell'Osservatorio 5,
  I-35122 Padova},altaddress={SISSA, Via Beirut 4, I-34014 Trieste}
}

\author{C. Burigana}{
  address={IASF-CNR, Via Gobetti, 101, I-40129 Bologna}
}

\author{A. Cavaliere}{
  address={Dipartimento Fisica, II Universit\`a di Roma,
  Via Ricerca Scientifica 1, 00133 Roma}
}

\author{L. Danese}{
  address={SISSA, Via Beirut 4, I-34014 Trieste}
}

\author{G.L. Granato}{
  address={INAF, Osservatorio Astronomico di Padova, Vicolo dell'Osservatorio 5,
  I-35122 Padova},altaddress={SISSA, Via Beirut 4, I-34014 Trieste}
}

\author{A. Lapi}{
  address={Dipartimento Fisica, II Universit\`a di Roma,
  Via Ricerca Scientifica 1, 00133 Roma}
}

\author{P. Platania}{
  address={Dipartimento di Fisica, Universit\`a di Milano, via
Celoria 16, I-20133 Milano}
}

\author{L. Silva}{
  address={INAF, Osservatorio Astronomico di Trieste, Via
G.B. Tiepolo 11, I-34131 Trieste}
}

\begin{abstract}
The Sunyaev-Zeldovich  (\cite{Sunyaev  Zeldovich 1972}; SZ) effect
has proven to be an extremely powerful tool to study the physical
and evolutionary properties of rich clusters of galaxies. Upcoming
SZ experiments, with their much improved sensitivity and angular
resolution, will provide unique information also on phases of
galaxy evolution characterized by the presence of large amounts of
hot proto-galactic gas. We present a preliminary analysis of the
SZ signals that can be expected at the collapse of the
proto-galaxy when, according to the standard scenario, the gas is
heated at its virial temperature, and during episodes of strong
energy injections from the active nucleus. The contributions of
such signals to excess power on arc-minute scales recently found
by CBI and BIMA experiments are briefly discussed.

\end{abstract}

\maketitle


\section{Introduction}

The currently standard hierarchical clustering paradigm for large
scale structure formation in a $\Lambda$CDM universe has
successfully confronted a broad variety of observations, ranging
from the formation of galaxy clusters, to large scale velocity
fields, to power spectra of the galaxy distribution and of the
microwave background, and more (see, e.g., \cite{Spergel et al.
2003}). However, the theory of formation and evolution of galaxies
is not in a very satisfactory state. Serious challenges have
emerged in the last years (\cite{Silk 2002}): the excess of
predicted small scale structure; the persistent inability of even
the best semi-analytic models (\cite{Devriendt  Guiderdoni
2000,Cole et al. 2000,Somerville et al. 2000,Menci et al. 2002})
to account for the surface density of massive galaxies at
substantial redshifts detected by (sub)-mm surveys with SCUBA and
MAMBO (\cite{Blain et al. 2002,Scott et al. 2002}) and by deep
K-band surveys (\cite{Cimatti e al. 2002,Kashikawa e al. 2002});
the difficulties to account for the distribution of velocity
dispersions of low-ionization damped Lyman-$\alpha$ systems at
$z>1.5$ (\cite{Prochaska  Wolfe 2001,Prochaska  Wolfe 2002}); the
low predicted specific angular momentum of galactic disks
(\cite{Navarro  Steinmetz 2000,Eke et al. 2001}); the
observational evidence contradicting the existence of the central
cusp in the dark matter distribution, predicted by numerical
simulations (\cite{Navarro et al. 1997,Moore et al. 1999,Klypin et
al. 2001}). On larger scales, the observed relationship between
X-ray luminosity and gas temperature for groups of galaxies
strongly deviates from expectations of the simplest hierarchical
clustering models.

Solutions of these problems may require advances in different
fields that may include particle physics (self-interacting dark
matter?), deviations from a power law of the power spectrum of
primordial density perturbations (as may be suggested by WMAP
data, \cite{Spergel et al. 2003}), or the astrophysics of galaxy
formation and evolution. In any case, a better understanding of
the complex physical processes governing the galaxy formation is
mandatory.

We discuss here how unique information may be provided by the
Sunyaev-Zeldovich (SZ) effect. In fact, the the proto-galactic gas
is expected to have a large thermal energy content, leading to a
detectable SZ signal, both when the protogalaxy collapses with the
gas shock-heated to the virial temperature (\cite{Rees
 Ostriker 1977,White  Rees 1978}), and in a later phase as
the result of strong feedback from a flaring active nucleus
(\cite{Ikeuchi 1981,Natarajan et al. 1998,Natarajan  Sigurdssson
1999,Aghanim et al. 2000,Platania et al. 2002,Lapi et al. 2003}).

\section{Galaxy-scale Sunyaev-Zeldovich effect}

Let us consider a fully ionized gas  with a thermal energy density
$\epsilon_{\rm gas}$, within the virial radius
\begin{equation} R_{\rm vir} = \left({3 M_{\rm vir}\over 4\pi
\rho_{\rm vir}}\right)^{1/3}\simeq 1.6\, 10^2 h^{-2/3} (1+z_{\rm
vir})^{-1} \left({M_{\rm vir}\over 10^{12} M_\odot}\right)^{1/3}\
\hbox{kpc},
\end{equation}
where $h$ is the Hubble constant in units of
$100\hbox{km}\,\hbox{s}^{-1}\,\hbox{Mpc}^{-1}$ and $\rho_{\rm
vir}\simeq 200 \rho_u$, $\rho_u=1.88\, 10^{-29}h^2(1+z_{\rm
vir})^3\,\hbox{g}\,\hbox{cm}^{-3}$ is the mean density of the
universe at the virialization redshift $z_{\rm vir}$.

The Comptonization parameter $y$, characterizing the amplitude of
the Sunyaev-Zeldovich effect due to this gas, can be estimated as
(\cite{Zeldovich  Sunyaev 1969}):
\begin{equation}
y \simeq {1\over 4} {\Delta \epsilon \over \epsilon_{\rm CMB}}\ ;
\label{y2}
\end{equation}
here $\epsilon_{\rm CMB} = a_{BB} T_{\rm CMB}^4 \simeq 4.2 \times
10^{-13}(1+z)^4\,\hbox{erg}\,\hbox{cm}^{-3}$, $a_{BB}$ being the
black-body constant and $T_{\rm CMB}= T_0(1+z)$ the temperature of
the cosmic microwave background (CMB). The amount $\Delta
\epsilon$ of gas thermal energy transferred to the CMB through
Thomson scattering is :
\begin{equation}
\Delta \epsilon \simeq  \epsilon_{\rm gas} {2(R_{\rm vir}/c)\over
t_c}\ , \label{Delta}
\end{equation}
$t_c$ being the gas cooling time by Thomson scattering.
%
%

A useful reference value for the thermal energy content of the
gas, $E_{\rm gas}$ is its binding energy ($E_{\rm b, gas}= M_{\rm
gas} v_{\rm vir}^2$, $v_{\rm vir}=162 h^{1/3}(1+z)^{1/2}(M_{\rm
vir}/10^{12}M_\odot)^{1/3}\,\hbox{km}\,\hbox{s}^{-1}$ being the
circular velocity of the galaxy at its virial radius, cf.
\cite{Navarro et al. 1997,Bullock et al. 2001}). The amplitude of
the SZ dip in the Rayleigh-Jeans region can be written as:
\begin{equation}
\left|\Delta T\right|_{\rm RJ} = 2yT_{\rm CMB} \simeq 1.7
\left({h\over 0.5}\right)^{2} \left({1+z_{\rm vir}\over
3.5}\right)^3 {M_{\rm gas}/M_{\rm vir}\over 0.1} {M_{\rm vir}\over
10^{12} M_\odot } {E_{\rm gas} \over E_{\rm b, gas}}\
\mu\hbox{K}\, .\label{DeltaT}
\end{equation}
Although current estimates of the ratio of the mass in stars to
the total mass in massive spheroidal galaxies yield average values
of $M_{\rm star}/M_{\rm vir}\simeq 0.03$ (\cite{McKay et al.
2001,Marinoni  Hudson 2002}), a value of $M_{\rm gas}/M_{\rm
vir}\simeq 0.1$, close to the cosmic ratio between baryon and
total mass density, is likely to be more appropriate for
proto-galaxies before a large fraction of the initial gas mass is
swept off by the combined action of supernova explosions and of
quasar feedback.

The SZ effect in Eq.~(\ref{DeltaT}) shows up on small (typically
sub-arcmin) angular scales. 
Quasar-driven blast-waves could inject into the ISM an amount of
energy several times higher than the gas binding energy, thus
producing larger, if much rarer, SZ signals.

\begin{figure}
  \includegraphics[height=.3\textheight,width=.7\textwidth]{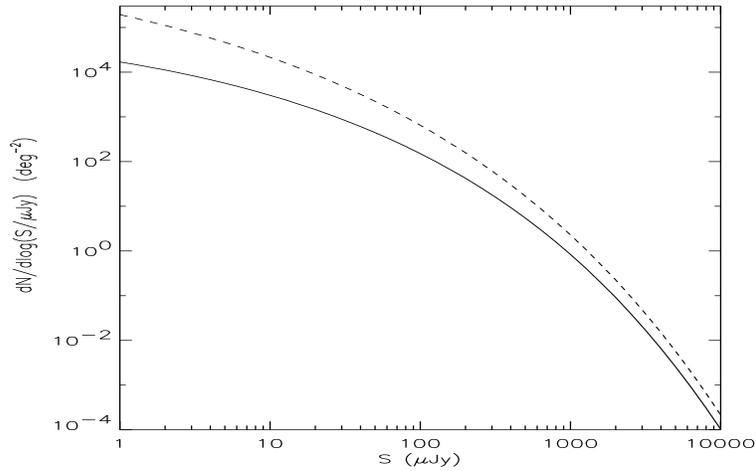}
  \caption{Counts of galaxy-scale SZ effects due to proto-galactic gas
  heated at the virial temperature (dashed curve) and to quasar-driven
  blast-waves (solid), at 30 GHz. The flux $S$ is the absolute value of
  $S_{\rm SZ}$ [eq.~(\ref{SSZ})]}
\end{figure}

\section{SZ effect from quasar feedback}

Impressively many lines of evidence converge in indicating that
the formation of super-massive black holes powering nuclear
activity is intimately linked to the formation of their host
galaxies (\cite{Ferrarese 2002,Yu  Tremaine 2002}. These include:

\begin{itemize}

\item the discovery that Massive Dark Objects (MDOs), with masses in
the range $\sim 10^6$--$3\times10^9\, M_\odot$ and a mass function
matching that of baryons accreted onto black holes during the
quasar activity (\cite{Salucci et al. 1999}), are present in
essentially all local galaxies with a substantial spheroidal
component (\cite{Kormendy  Richstone 1995,Magorrian et al.
1998,van der Marel 1999,Kormendy  Gebhardt 2001});

\item the tight correlation between the MDO mass (M$_{\rm BH}$)
and the velocity dispersion ($\sigma$) of stars in the host galaxy
(\cite{Magorrian et al. 1998,Ferrarese  Merritt 2000,Gebhardt et
al. 2000,Tremaine et al. 2002}), the mass of the spheroidal
component (\cite{McLure  Dunlop 2002,Dunlop et al. 2003}), and the
mass of the dark halo \cite{Ferrarese 2002}; recently
\cite{Shields et al. (2003)} have found that the M$_{\rm
BH}$--$\sigma$ correlation is already present at redshift up to
$\sim 3$;
\item the correspondence between the luminosity function of active
star-forming galaxies at $z\simeq 3$, the B-band luminosity
function of quasars, and the mass function of dark halos at the
same redshift (\cite{Haehnelt et al. 1998,Monaco et al. 2000});

\item the similarity between the evolutionary histories of the
luminosity densities of galaxies and quasars (e.g. \cite{Cavaliere
Vittorini 1998}).

\end{itemize}

\noindent As a consequence, the evolutionary histories of both
active nuclei and spheroidal galaxies (or galactic bulges) can
only be understood if we properly allow for their mutual feedback.
As discussed by \cite{Granato et al. 2001,Granato et al. 2003}),
the mutual feedback between galaxies and quasars during their
early evolutionary stages may indeed be the key to overcome some
of the crises of the currently standard scenario for galaxy
evolution.

One ingredient of the scenario proposed by \cite{Granato et al.
2003} are powerful quasar driven outflows, carrying a considerable
fraction of the quasar bolometric luminosity, $L_{\rm bol}$,
increasing as $L_{\rm bol}^{1/2}$ (for emission close to the
Eddington limit). For large enough galaxies ($M_{\rm vir} {\ge}
10^{12}\,M_\odot$), the interstellar medium (ISM) is eventually
swept out, and, correspondingly, the star-formation is switched
off, on shorter timescales for more massive objects. The outflows
are expected to be highly supersonic, and therefore liable to
induce strong shocks, transiently heating the interstellar gas to
temperatures exceeding the virial value.

A black-hole (BH) accreting a mass $M_{\rm BH}$ with a mass to
radiation conversion efficiency $\epsilon_{\rm BH}$ releases an
energy $E_{\rm BH}=\epsilon_{\rm BH}M_{\rm BH}c^2$. We adopt the
standard value for the efficiency $\epsilon_{\rm BH}=0.1$ and
assume that a fraction $f_h=0.1$ of the energy is fed in kinetic
form and generates strong shocks turning it into heat. Actually
the fraction, $f_h$, of energy going into heating of the
interstellar gas may be a function of the BH mass, or of the
bolometric luminosity of the active nucleus (AGN). For example,
\cite{Granato et al. 2003} argue that, based on the model for
AGN-driven outflows by \cite{Murray et al. 1995}, for emission at
the Eddington limit, the fraction of bolometric luminosity
released in kinetic form increases as $L_{\rm bol}^{1/2}$. This
conclusion is, however, model-dependent and we prefer the simpler
assumption of a constant value for $f_h$, that may be viewed as an
effective, luminosity weighted value. As discussed by
\cite{Cavaliere et al. 2002,Platania et al. 2002}, $f_h\sim 0.1$
can account for the pre-heating of the intergalactic medium in
groups of galaxies.

Using the recent re-assessment by \cite{Tremaine et al. 2002} of
the well known correlation between the BH mass and the stellar
velocity dispersion
\begin{equation}
M_{\rm BH} = 1.4\, 10^8\,\left( {\sigma \over 200\,{\rm
km/s}}\right)^{4}\ {\rm M}_\odot \ . \label{Mbh}
\end{equation}
we get
\begin{equation}
{E_{\rm BH}\over E_{\rm b, gas}}\simeq 4.7  \left({h\over
0.5}\right)^{-2/3}{\epsilon_{\rm BH}\over 0.1}\, {f_h\over 0.1} \,
{1+z_{\rm vir}\over 3.5}\, \left({M_{\rm gas}/M_{\rm vir}\over
0.1}\right)^{-1} \left({M_{\rm vir}\over 10^{12}
M_\odot}\right)^{-1/3}\ . \label{ratioT}
\end{equation}
%
%
%
The amplitude of the SZ dip in the Rayleigh-Jeans region due to
quasar heating of the gas is then estimated as:
\begin{equation}
\left|\left({\Delta T \over T}\right)_{\rm RJ}\right| \simeq
1.8\times 10^{-5} {f_h \over 0.1} \left({h\over 0.5}\right)^2
\left({\epsilon_{\rm BH} \over 0.1}\right)^{1/2} \left({E_{\rm
BH}\over 10^{62}}\right)^{1/2}\left({1+z \over 3.5}\right)^{3}\
.\label{DeltaTqso}
\end{equation}
Following \cite{Platania et al. 2002}, we adopt an isothermal
density profile of the galaxy. The virial radius, encompassing a
mean density of $200\rho_u$, is then:
\begin{equation}
R_{\rm vir} \simeq  120 \left({h\over 0.5}\right)^{-1}
\left({E_{\rm BH}\over 10^{62}}\right)^{1/4} \left({\epsilon_{\rm
BH}\over 0.1}\right)^{-1/4}\left(1+z_{\rm vir} \over
3.5\right)^{-3/2}\ \hbox{kpc}\ , \label{Rg}
\end{equation}
corresponding to an angular radius:
\begin{eqnarray}
\theta_{SZ}\simeq 17'' \left({E_{\rm BH}\over
10^{62}}\right)^{1/4} \left({\epsilon_{\rm BH}\over
0.1}\right)^{-1/4}\left(1+z_{\rm vir} \over
3.5\right)^{-3/2}{d_A(2.5)\over d_A(z)}\ , \label{theta}
\end{eqnarray}
where $d_A(z)$ is the angular diameter distance.

\begin{figure}
  \includegraphics[height=.3\textheight,width=.7\textwidth]{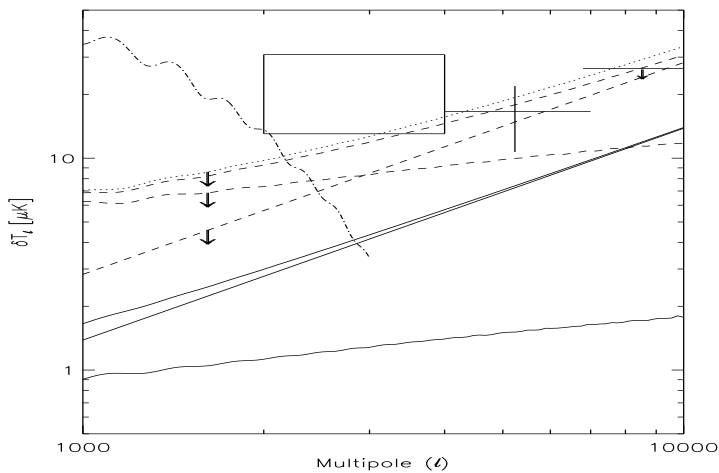}
  \caption{Power spectra of galaxy-scale SZ effects at 30 GHz, compared
  to the CMB power spectrum and to the
  CBI (\cite{Mason et al. 2003}; large square) and BIMA (\cite{Dawson et al. 2002};
  point with 68\% confidence error bars and 95\% confidence upper limit)
  data. The dashed lines correspond to
  SZ effects due to proto-galactic gas, the solid lines to quasar-driven
  blast-waves. For each set, the steep straight line refer to Poisson fluctuations,
  the flatter line to fluctuations due to clustering, the uppermost
  line to the sum (in quadrature) of the two contributions. The dotted line
  is the sum (in quadrature) of all contributions. The downward arrows
  signify that, as mentioned in the text, the contributions of proto-galactic
  gas (dashed lines) are actually upper limits. We also argue that the
  CBI results (large square) might be, conservatively, interpreted as
  upper limits}
\end{figure}

\section{Counts and small-scale fluctuations}

The "flux density", which in the Rayleigh-Jeans region is actually
negative, associated to the SZ effect is given by (neglecting
relativistic corrections):
\begin{equation}
S_{\rm SZ}(\nu)={2\left(kT_{\rm CMB}\right)^3\over
(hc)^2}g(x)y\omega\ , \label{SSZ}
\end{equation}
where $\omega = \pi \theta_{\rm SZ}^2$ and
\begin{equation}
g(x)={x^4e^x\over\left(e^x-1\right)^2}\left(x{e^x+1\over
e^x-1}-4\right)\ , \label{gx}
\end{equation}
with $x=h\nu/kT_{\rm CMB}$.

\subsection{Quasar-driven SZ signals}

The counts  of quasar-driven SZ signals, and the corresponding
small scale fluctuations, can be estimated from the evolving
B-band luminosity function of quasars, $\phi(L_B,z)$, relating the
total amount of energy released, $E_{\rm BH}$, to the B-band
luminosity, $L_B$. If the quasar radiates at the Eddington limit
$E_{\rm BH} =k_{\rm bol}L_Bt_S$, where $k_{\rm bol}$ is the
bolometric correction and $t_S\simeq 4.5\times 10^7\,\hbox{yr}$ is
the quasar lifetime for $\epsilon_{\rm BH}=0.1$.

The lifetime, $t_{\rm SZ}$, of the transient SZ effect due to the
quasar-driven blast-waves is approximately the time for the shock
to reach the outer border of the host galaxy. Using the expression
(\cite{Ostriker  McKee 1988}) for the evolution with time of the
radius of a self-similar blast-wave expanding in a medium with an
isothermal density profile, $\rho \propto r^{-2}$, we have:
\begin{equation}
t_{\rm SZ}\simeq 1.5\times 10^8 \left({h\over 0.5}\right)^{-3/2}
\left({E_{\rm BH} \over
10^{62}\hbox{erg}}\right)^{1/8}\left({\epsilon_{\rm BH}\over
0.1}\right)^{-5/8}\left({f_h\over
0.1}\right)^{-1/2}\left({1+z\over 3.5}\right)^{-9/4}\ \hbox{yr}.
\label{tSZ}
\end{equation}
The number density of SZ sources per unit interval of SZ ``flux''
$S_{\rm SZ}$ can then be estimated as
\begin{equation}
\phi_{\rm SZ}(S_{\rm SZ},z) = \phi(L_B,z) {t_{\rm SZ} \over t_{\rm
q,opt}}\,{d L_B \over d S_{\rm SZ}} \ , \label{phiSZ}
\end{equation}
where $L_B(S_{\rm SZ},z)$ is the blue luminosity of a quasar at
redshift $z$ causing an SZ flux $S_{\rm SZ}$, and $t_{\rm q,opt}$
is the duration of the optically bright phase of the quasar
evolution.

Figure~1 shows, as an example, the predicted counts at 30 GHz, as
a function of $S_{\rm SZ}$, obtained adopting the exponential
model for the evolving luminosity function of quasars derived by
\cite{Pei 1995} for an Einstein-de Sitter universe, an Hubble
constant of $50\,\hbox{km}\,\hbox{s}^{-1}\,\hbox{Mpc}^{-1}$, and
an optical spectral index of quasars $\alpha=0.5$ ($S_\nu \propto
\nu^\alpha$). The parameters have been set at $\epsilon_{\rm
BH}=0.1$, $f_h=0.1$, $k_{\rm bol}=10$, $t_{\rm
q,opt}=10^7\,\hbox{yr}$.

\subsection{SZ signals from proto-galactic gas at virial
temperature}

As discussed by \cite{Granato et al. 2001} quasars can be used as
effective signposts for massive spheroidal galaxies in their early
evolutionary phases; this allows us to bypass the uncertainties in
the normalization of the primordial perturbation spectrum
affecting estimates based on the various versions of the Press \&
Schechter formalism. According to \cite{Ferrarese 2003}, the mass
of their dark-matter halo, $M_{\rm vir}$, is related to the mass
of the central black-hole by:
\begin{equation}
{M_{\rm BH}\over 10^8 M_\odot}\sim 0.1 \left({M_{\rm vir} \over
10^{12} M_\odot}\right)^{1.65} \ . \label{Ferr}
\end{equation}
Under the assumption that quasars emit at the Eddington limit, the
number density of sources with gas at virial temperature,
producing SZ dips of amplitude given by Eq.~(\ref{DeltaT}), can be
straightforwardly related to the quasar luminosity function
$\phi(L_B,z)$ by an equation analog to Eq.~(\ref{phiSZ}),
replacing $t_{\rm SZ}$ with the gas cooling time, $t_{\rm cool}$.

The counts shown in Fig.~1 refer to the same cosmological model
and the same lifetime of the quasar optically bright phase as
above and to $M_{\rm gas}/M_{\rm vir} = 0.1$. As for the cooling
time, which is rather uncertain, we have made the extreme
assumption that $t_{\rm cool}=t_{\rm exp}$. Thus the counts
plotted have to be taken as upper limits.

\subsection{Small scale fluctuations}

The Poisson fluctuations are straightforwardly computed from the
counts; since the latter are very steep, the results are
insensitive to the upper flux density cutoff.

To estimate fluctuations due to clustering we adopted the spatial
correlation function of quasars estimated by \cite{Croom et al.
2002} for an Einstein-de Sitter universe
($\xi(r)=(r/r_0)^{-1.58}$, with $r_0=4.29h^{-1}\,$Mpc, constant in
comoving coordinates), cut-off at $r=3r_0$.

The power spectra of fluctuations due to both the gas in virial
equilibrium and heated by the quasar feedback are compared, in
Fig.~2, with the recent measurements of arcminute scale
fluctuations at 30 GHz by the CBI (\cite{Mason et al. 2003}) and
BIMA (\cite{Dawson et al. 2002}) experiments. We stress again that
our estimated of signal from the gas at the virial temperature is
based on an over-simplified treatment. In fact, the thermal
history of the protogalactic gas is quite complex: only a fraction
of it may be heated to the virial temperature (\cite{Binney
2003,Birnboim Dekel 2003}); cooling may be relatively rapid in the
densest regions; on the other hand, significant heating may be
provided by supernovae and by the central active nucleus. Also, we
expect that (\cite{Granato et al. 2003}), particularly in the case
of the most massive halos virializing at high $z$, the quasar
feedback removes the residual interstellar gas on a timescale
ranging to 0.5 to a few Gyrs. Thus, again, the results shown in
Fig.~2 are (possibly generous) upper limits.

\section{Discussion and conclusions}

Evidences of statistically significant detections at 30 GHz of
arcminute scale fluctuations in excess of predictions for
primordial anisotropies of the cosmic microwave background (CMB)
and also in excess of the estimated contamination by extragalactic
radio sources have recently been obtained by the CBI (\cite{Mason
et al. 2003}) and BIMA (\cite{Dawson et al. 2002}) experiments.

It may be noted that, although the CBI group applied a quite
elaborated treatment of radio-sources, the sensitivity of their 31
GHz point source observations does not really guarantee proper
allowance for this foreground. Their strategy comprised pointed 31
GHz observations with a $4\sigma$ detection limit of $S_{31 {\rm
GHz}} = 6\,$mJy, of all NVSS sources with $S_{1.4 {\rm GHz}} \geq
6\,$mJy and a direct survey reaching a flux limit of $S_{{\rm
lim}, 31 {\rm GHz}} = 6\,$mJy, over an area of 1.8 square degrees.
If the slope of their 31 GHz counts keeps constant down to flux
densities several times lower than $S_{{\rm lim}, 31 {\rm GHz}}$,
the Poisson fluctuations due to sources below the detection limit
amount to $\simeq 30\,\mu$K, and may entirely account for the
measured signal. Thus, their excess signal (15-$30\,\mu$K in the
multipole range $\ell =2000$--4000) may be more conservatively
interpreted as an upper limit.

The BIMA group (\cite{Dawson et al. 2002}) carried out a VLA
survey at 4.8 GHz of their fields down to a flux density limit of
$\sim 150\,\mu$Jy to identify and remove point sources. In this
case, the residual contamination is indeed likely to be small, at
the cost of a loss of sensitivity.

If indeed the detected signal cannot be attributed to
extragalactic radio sources, its most likely source is the thermal
SZ effect (\cite{Gnedin Jaffe 2001}). The SZ within rich clusters
of galaxies has been extensively investigated (\cite{Komatsu
Kitayama 1999,Bond et al. 2003}). The estimated power spectrum was
found to be very sensitive to the normalization ($\sigma_8$) of
the density perturbation spectrum. A normalization $\sigma_8\ge
1$, somewhat higher than indicated by other data sets, is
apparently required to account for the CBI data. High resolution
hydrodynamical simulations of structure formation (\cite{Zhang et
al. 2002,White et al. 2002}) have highlighted the presence of
substantial SZ structure on sub-arcmin scales.

We have carried out an analytical investigation of the SZ signals
associated to the formation of spheroidal galaxies. We find that
they are potentially able to account for the BIMA results.
Proto-galactic gas heated at the virial temperature and with a
cooling time comparable with the expansion time may provide the
dominant contribution; in this case we expect clustering
fluctuations of amplitude comparable to Poisson fluctuations,
although with a flatter power spectrum.

Potentially detectable SZ signals can also be produced by strong
feedback from the central active nucleus. It is, in fact, becoming
increasingly clear that the interplay between star-formation and
nuclear activity plays a key role in shaping the early evolution
of both the host galaxy and the active nucleus. The enormous
amounts of energy that the active nuclei (AGNs) release to their
environment, not only in radiative but also in kinetic form, can
strongly influence the gas both in the host galaxy and in the
surrounding intergalactic medium, with conspicuous effects on
X-ray emission and on the SZ signal. A flaring AGN can drive an
energetic shock, causing a transient, strong rise of the average
gas pressure and, consequently, of the amplitude of the SZ effect.
These signals are much rarer but can individually reach
substantially higher values.

High frequency surveys with sub-arcmin resolution can directly
detect individual SZ sources. We expect a surface density of
$\simeq 1\,\hbox{deg}^{-2}$ at $S_{30{\rm GHz}}\simeq 1\,$mJy. Of
course, the counts at these relatively bright fluxes are likely
dominated by SZ effects in clusters of galaxies, which however
should be easily distinguishable because of their much larger
angular size and much lower redshift.

We thus conclude that the SZ effect is an effective probe of the
thermal state of the gas, of its evolution and, in particular, of
the AGN feedback. The main parameter governing the global effect
on the ambient medium of the energy released by AGNs is the ratio,
$\Delta E/E$, between the amount of energy deposited in the medium
and the binding energy of the receiving gas (\cite{Cavaliere et
al. 2002}). $\Delta E/E$ can plausibly be $\geq 1$ for very
massive galaxies and small groups, but is $\ll 1$ for rich
clusters. Qualitatively different effects are therefore expected
in the two cases. If ${\Delta E/E} {\ge} 1$ the energy injection
heats the gas and partially ejects it out of the potential well.
This may lead directly to a correlation between the black-hole
mass and the velocity dispersion of stars (or the bulge mass) with
slope and normalization in nice agreement with the data. Also, the
ensuing increase of the gas temperature and decrease of the X-ray
luminosity (because of the decreased gas density) may explain the
deviation of the observed $T$--$L_X$ from predictions of the
hierarchical clustering scenario, in the presence of gravitational
effects only. Ongoing and forthcoming high resolution surveys in
the cm/mm region may directly test this scenario. Clearly the SZ
effect is a much more effective probe of the thermal energy
content of the plasma than the X-ray emission. The former is in
fact an almost perfect calorimeter (\cite{Birkinshaw 1999}), being
directly proportional to the thermal energy. On the contrary,
X-ray emission is very sensitive to the details of the density
distribution of the emitting gas.

\begin{theacknowledgments}
Work supported in part by ASI and MIUR.
\end{theacknowledgments}


\bibliographystyle{aipprocl} 



\end{document}